\documentclass[letter,scriptaddress,aps,noshowkeys,noshowpacs,superscriptaddress]{revtex4}

	\usepackage{natbib}
	\usepackage{amsmath}
	\usepackage{makeidx}
	\usepackage{amsfonts}
	\usepackage[ansinew]{inputenc}
	\usepackage[usenames,dvipsnames]{pstricks}
	\usepackage{subfigure}
	\usepackage{epsfig}
	\usepackage{pst-grad} 
	\usepackage{pst-plot} 
	\usepackage[colorlinks,hyperindex]{hyperref}
	\usepackage{epstopdf}
	\usepackage{multirow}
	\usepackage{titlesec}
	\usepackage{float}
	\usepackage{amssymb}
	\usepackage{siunitx}

	\titlespacing{\section}{0pt}{5pt}{5pt}
	\titlespacing{\subsection}{0pt}{5pt}{5pt}
	\titlespacing{\subsubsection}{0pt}{5pt}{5pt}

	\hypersetup
	{
		colorlinks,%
		citecolor=black,%
		linkcolor=black,%
		urlcolor=black,%
	}

\makeindex

\begin{document}

\title{Generation and analysis of correlated pairs of photons on board a nanosatellite}

\author{Zhongkan Tang}
\affiliation{Centre for Quantum Technologies, National University of Singapore,\\ Block S15, 3 Science Drive 2, 117543 Singapore.}
\author{Rakhitha Chandrasekara}
\affiliation{Centre for Quantum Technologies, National University of Singapore,\\ Block S15, 3 Science Drive 2, 117543 Singapore.}
\author{Yue Chuan Tan}
\affiliation{Centre for Quantum Technologies, National University of Singapore,\\ Block S15, 3 Science Drive 2, 117543 Singapore.}
\author{Cliff Cheng}
\affiliation{Centre for Quantum Technologies, National University of Singapore,\\ Block S15, 3 Science Drive 2, 117543 Singapore.}
\author{Luo Sha}
\affiliation{Department of Electrical and Computer Engineering, National University of Singapore,\\ Block E4, 4 Engineering Drive 3, 117583 Singapore.}
\author{Goh Cher Hiang}
\affiliation{Department of Electrical and Computer Engineering, National University of Singapore,\\ Block E4, 4 Engineering Drive 3, 117583 Singapore.}
\author{Daniel K. L. Oi}
\affiliation{SUPA Department of Physics, University of Strathclyde\\ John Anderson Building, 107 Rottenrow East, G4 0NG Glasgow, UK}
\author{Alexander Ling}
\affiliation{Centre for Quantum Technologies, National University of Singapore,\\ Block S15, 3 Science Drive 2, 117543 Singapore.}
\affiliation{Department of Physics, National University of Singapore,\\ 3 Science Drive 2,  117551 Singapore}


\begin{abstract}
Satellites carrying sources of entangled photons could establish a global quantum network, enabling private encryption keys between any two points on Earth.
Despite numerous proposals, demonstration of space-based quantum systems has been limited due to the cost of traditional satellites.  We are using very small spacecraft to accelerate progress. We report the in-orbit operation of a photon pair source aboard a \SI{1.65}{\kg} nanosatellite and demonstrate pair generation and polarization correlation under space conditions. The in-orbit photon correlations exhibit a contrast of $97\pm2$\%, matching ground-based tests. This pathfinding mission 
overcomes the challenge of demonstrating in-orbit performance for the components of future entangled photon experiments. Ongoing operation establishes the in-orbit lifetime of these critical components. More generally, this demonstrates the ability for nanosatellites to enable faster progress in space-based research.
\end{abstract}


\maketitle

Progress in quantum computers and their threat to public key cryptography is driving new forms of encryption~\cite{cnsa}. One promising alternative is quantum key distribution (QKD) which provides provable security underpinned by quantum physics~\cite{gisin02}.
In particular, QKD can be achieved using pairs of photons that possess fundamental correlations known as quantum entanglement~\cite{ekert91}. Practically, entanglement-based QKD enables a reduction in the number of trusted components~\cite{acin07}. 
A global quantum network for distributing entangled photon pairs will enable strong encryption keys to be shared between any two points on Earth.

Entanglement-based QKD is a mature technology~\cite{jennewein00,ursin07,ling08_1} compared with other entanglement-assisted applications. However, it shares a common range limit with more conventional prepare-send-measure QKD schemes~\cite{mosca12}. Metropolitan scale QKD networks are possible using optical fiber or free-space links, but fiber losses and ground-level atmospheric turbulence preclude extending these networks to a global scale. Quantum repeater technology that can overcome these losses is still in the starting stages of being researched~\cite{kimble08}. 

Scalable global entanglement distribution can be achieved using a constellation of satellites equipped with space-to-ground optical links~\cite{boone15}. The technology behind these links are relatively well-documented (e.g.~\cite{elser15,vallone15,opals}). Most proposals~\cite{jennewein13} employ a downlink to minimize transmission loss~\cite{bourgoin13}. A source of photon pairs is placed on board a satellite in Earth orbit that will then either act as a trusted relay between two ground nodes, or beam one photon to one ground station and its pair photon to a different ground station.
An additional advantage of space-based entanglement systems is that they allow fundamental tests of the possible overlap between quantum and relativistic regimes for which operation in space is necessary~\cite{rideout12}. 

Despite many preliminary studies on space quantum systems~\cite{buttler00, kurtsiefer02, rarity02, wang13, bourgoin13, elser15, tang14, vallone15} there has been limited published work demonstrating relevant technology in space~\cite{tang14, vallone15} due to the prohibitive cost of traditional space mission development that has hindered in-situ experimental progress.
In particular, very little work has been done on adapting bright and efficient (but bulky and delicate) laboratory-based photon pair sources into space-qualified systems.
Currently there is one satellite that has been announced that will carry an entangled photon source~\cite{horiuchi15}. 
Greater access to space will enable more in-orbit experiments leading to a greater rate of experimentation and innovation.

\section*{A new pathway to space}
\label{nanosat}
The emergence of very small spacecraft (below \SI{10}{\kg} in mass) called nanosatellites has made access to space more cost-effective. The capability of nanosatellites has rapidly improved due to the development of miniaturized systems capitalizing on commercial-off-the-shelf components. The most common nanosatellite standard is the cubesat~\cite{woellert11}; a \SI{10}{\cm} cube in its smallest configuration, modularly expandable into cuboid spacecraft. Nanosatellites are typically launched piggy-back with conventional large spacecraft, keeping launch costs low.

The relatively low cost of nanosatellite missions enables shorter development cycles, the use of non-space rated components, and an iterated approach to space mission development. Through multiple launches, the space worthiness of components can be established, leading to greater confidence that critical sub-systems will perform well in the final mission, an approach that has also been adopted by larger space missions~\cite{lisa}.

We have adopted this iterative approach to the demonstration of a bright polarization-entangled photon pair source in space. The first step is to address the technical challenges of assembling a photon pair source that satisfies the restrictions imposed by very small spacecraft. Though the nanosatellite platform imposes strict size, weight and power constraints on payloads, this is used to drive the development of compact and rugged photon pair systems~\cite{tang15} that will also be useful for ground-based applications of entangled photons.

In this paper, we report the use of a nanosatellite to conduct an in-orbit photon counting experiment where the polarization correlation between pairs of photons was measured. The demonstration serves two objectives. The first objective is to develop the techniques for assembling the optics necessary for a space-capable photon pair source that could, in the future, be rapidly converted to produce polarization-entangled photon pairs. The second objective is to use the generated photon pairs for metrology to establish the space worthiness of the apparatus used to detect and measure the polarization correlations.

\section*{A small photon pair source}
\label{payload}
The workhorse technique for generation of entangled pairs of photons relies on a quantum process called spontaneous parametric downconversion (SPDC)~\cite{coinc70}. In SPDC, a parent photon within an optical nonlinear material is converted into a pair of daughter photons obeying energy and momentum conservation. These conservation laws result in the daughter photons exhibiting classical correlations in a number of physical properties such as energy, time and polarization. These classical correlations are a common building block in a large number of methods used in experimental optics~\cite{hong87} and also of practical interest in metrology~\cite{migdall01} or time synchronisation~\cite{ho09}. By careful arrangement of the SPDC process, the photon pair can be put into a suitable superposition and become entangled~\cite{shih88}. 

We chose a photon pair source based on non-degenerate, collinear Type-I SPDC built around a single $\beta$-Barium Borate (BBO) material as this design satisfies the stringent size and weight constraints in a nanosatellite. This design can be built to achieve the final brightness and efficiency (pair-to-singles ratio) in single-mode optics that are needed in a space-to-ground QKD link. A source emitting useful photon pairs in a collinear direction is compact, and a temperature tolerant material such as BBO reduces the power requirements for thermal control. Together, these result in considerable weight savings. This geometry can be readily extended into a high quality entangled photon pair source~\cite{trojek08} by using additional non-linear crystals.

\begin{figure}[t]
\center
\includegraphics[width=0.75\linewidth]{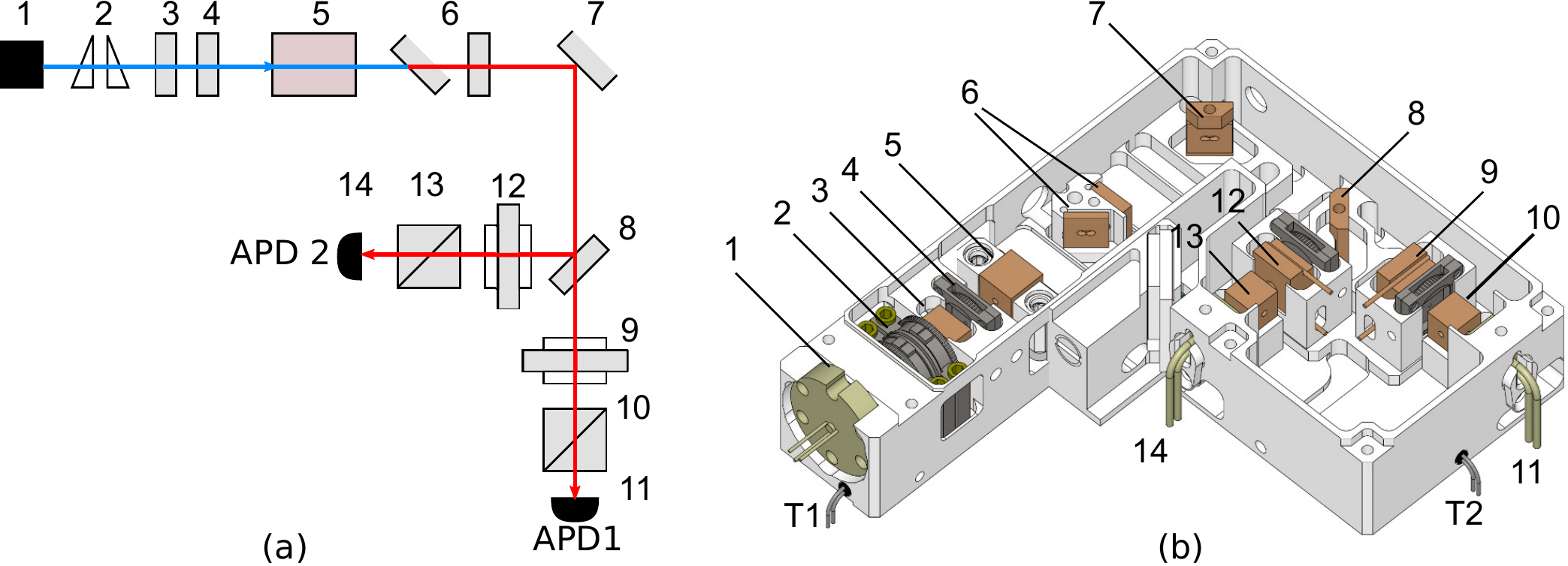}    
\caption{\label{fig:source} (a) Schema of components in the photon pair source (items 1 to 6) and the measurement section (7-14). A \SI{405}{\nm} pump diode~(1) supplies the pump beam aligned by a pair of prisms~(2). The pump is prepared by a band-pass filter~(3) and polarization half-wave plate~(4) to maximize SPDC production within the BBO crystal~(5). Excess pump beam is removed from the optical path using filters~(6). A gold mirror~(7) folds the optical path. Signal and idler photons are separated by a dichroic mirror~(8) before the photons are interrogated by polarisation rotators~(9 \& 12) and polarizing beam splitters~(10 \& 13) in front of Geiger-mode avalanche photodiodes, APD1~(11) \& APD2~(14). (b) The 3-D model of the optics. The crystal is mounted on a steel flexure stage that enables the crystal tilt to be adjusted by screws. Also shown is the position of temperature sensors T1 and T2. A thin-film heater (not shown) for raising the payload temperature is located below APD2. The optical housing has sufficient volume for additional crystals that can be inserted to enable the generation of polarization-entangled photon pairs.}
\end{figure}

The source layout is shown in Fig.~\ref{fig:source}. One of the most demanding challenges was to ensure that the BBO crystal axis made an angle of \SI{28.8}{\degree} to the pump beam to within a tolerance of $\pm$\SI{100}{\micro\radian}. This precision is typically achieved in a laboratory using bulky kinematic mounts unsuitable for use within a nanosatellite due to their size and sensitivity to mechanical shock and vibration. A custom steel flexure stage was designed to mount the crystal, enabling the correct angle and tolerance to be achieved. Pump light at \SI{405}{\nm} generates signal (\SI{760}{\nm}) and idler (\SI{867}{\nm}) photons in a \SI{6}{\mm} length crystal. The full-width at half-maximum bandwidth of the signal and idler photons are each approximately \SI{17}{\nm}. 

After separation by a dichroic mirror, the individual signal and idler photons pass through a polarization rotator and filter (the transmission arm of a polarizing beamsplitter whose extinction ratio is 200:1). The rotators are implemented using nematic liquid crystals. The lack of moving parts avoids de-stabilizing the spacecraft. The single photons are detected by Geiger-mode avalanche photodiodes (GM-APDs) operated using a closed-loop control circuit~\cite{cheng15}. Correlation data is acquired by setting the polarization rotator for the idler photons at a fixed voltage setting, while the rotator for the signal photons is stepped through a series of pre-determined voltages corresponding to different amounts of polarization rotation. Ground-based tests have established that the rotators can achieve 2$\pi$ rotation with a polarization contrast of 97\% over a wide temperature range. The value of the contrast is limited by the rotator's intrinsic performance. 

Amongst the components flown in this pathfinder experiment, the rotators and the GM-APDs are of special interest as in-orbit radiation could rapidly lead to permanent damage. Damaged liquid crystals exhibit a poor polarization rotation ability. For the photodiodes, radiation damage leads to an increased rate of noise events (also known as dark counts) that ultimately drowns out the optical signal. The photon pairs generated using the SPDC process will be used to test the in-orbit performance of both components. In particular, the full performance of the polarization rotator can be re-constructed by observing their effect on the strongly co-polarized SPDC photons using Malus' law. The rotators and detectors had previously exhibited good performance in radiation tests~\cite{tan13,tan15}, and simulation studies of the in-orbit radiation environment predict that the devices should perform beyond the spacecraft useful life~\cite{tan13}. This will be checked by periodic observation of the orbiting source and comparison with a ground-based copy.

The optical components are housed within a light-tight aluminium box. The power needed to run the optical payload is \SI{1.3}{\W}, two orders of magnitude smaller than needed in a comparable laboratory setup. The complete payload mass is \SI{220}{\g}. The physical footprint of the photon pair source (approximately 10~cm$\times$10~cm) is significantly smaller than most SPDC experiments that use optical tables to accommodate the pump laser, crystal adjusters and photon detectors. The design retains sufficient spare volume for additional crystals to be inserted that enable the generation of polarization-entangled photon pairs. 

Two copies of the optics payload were assembled and programmed with an identical set of experiment profiles (see Methods) which can be selected by ground control. Each experiment profile operates for a maximum duration of 30 minutes to avoid straining the spacecraft power system. One copy successfully underwent a full qualification test process (involving vibration, and thermal-vaccuum conditions)  to validate the robustness of the design. The second copy was integrated into the spacecraft which subsequently passed the acceptance test regime set by the launch vehicle provider~\cite{tang15}. Performance data of the flight copy whilst integrated into the spacecraft, to act as a comparative baseline, was taken before shipment of the spacecraft to the launch site. We note that it was not possible to exactly replicate the expected in-orbit temperature environment for this final baseline. The elevated pre-launch operations temperature led to an undesired deflection (approximately \SI{170}{\micro\radian}) in the steel flexure stage causing misalignment of the BBO cyrstal, reducing the effectiveness of the SPDC process in this final baseline test. This deflection is temperature dependent, and the crystal was expected to be correctly aligned when the satellite was deployed.

\section*{In-orbit operation and performance comparison}
\label{spacedemo}
The host spacecraft (Galassia)~\cite{galassia} was placed in orbit at an altitude of approximately \SI{550}{\km} with an inclination of \SI{15}{\degree}. In this orbit, the spacecraft makes multiple daily passes over the Singapore-based control station. 
The first opportunity to activate the payload using the default settings occurred after 36 days in orbit, by which time the spacecraft had already experienced over 500 diurnal cycles. During a single orbit the spacecraft internal temperature oscillates between -\SI{2}{\celsius} and \SI{26}{\celsius}. 

\begin{figure}[t]
\center
\includegraphics[width=0.8\linewidth]{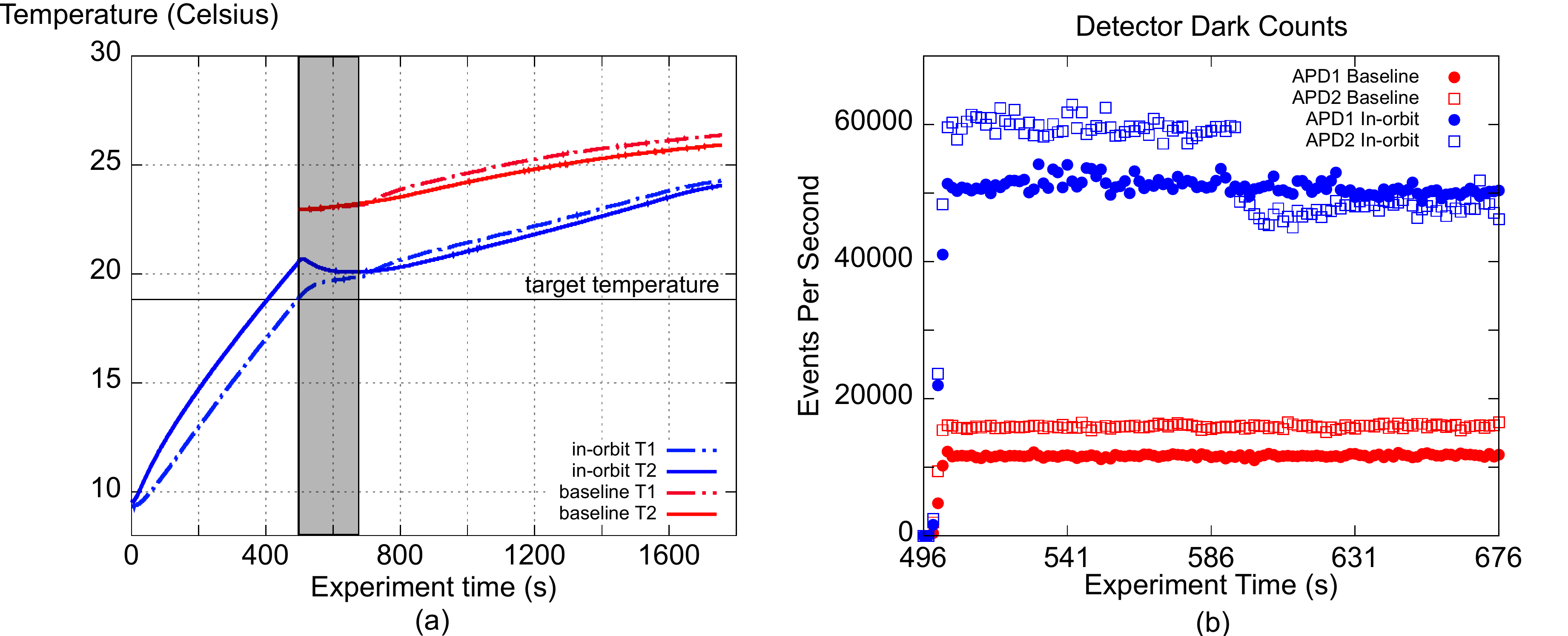}    
\caption{\label{fig:data} (a) Payload temperatures recorded during operation in space (blue). For comparison, the temperature from the final test on Earth is provided (red). At activation (t=0), the payload was heated to the target temperature (\SI{18}{\celsius}). The detectors were then run for three minutes to observe their noise performance (grey region), following which polarization measurement commenced. The T2 sensor initially recorded a higher reading due to its proximity to the heater. The Earth-based test started at a higher temperature and did not involve heater operation. (b) Comparison of detector noise (dark counts) observed on Earth and in-orbit. The in-orbit dark count rate is higher than predicted by models that simulate radiation damage. The drop in dark count rate for APD2 is attributed to a fall in temperature after the heater (located beneath APD2) is turned off.  }
\end{figure}

\begin{figure}[]
\center
\includegraphics[width=0.5\linewidth]{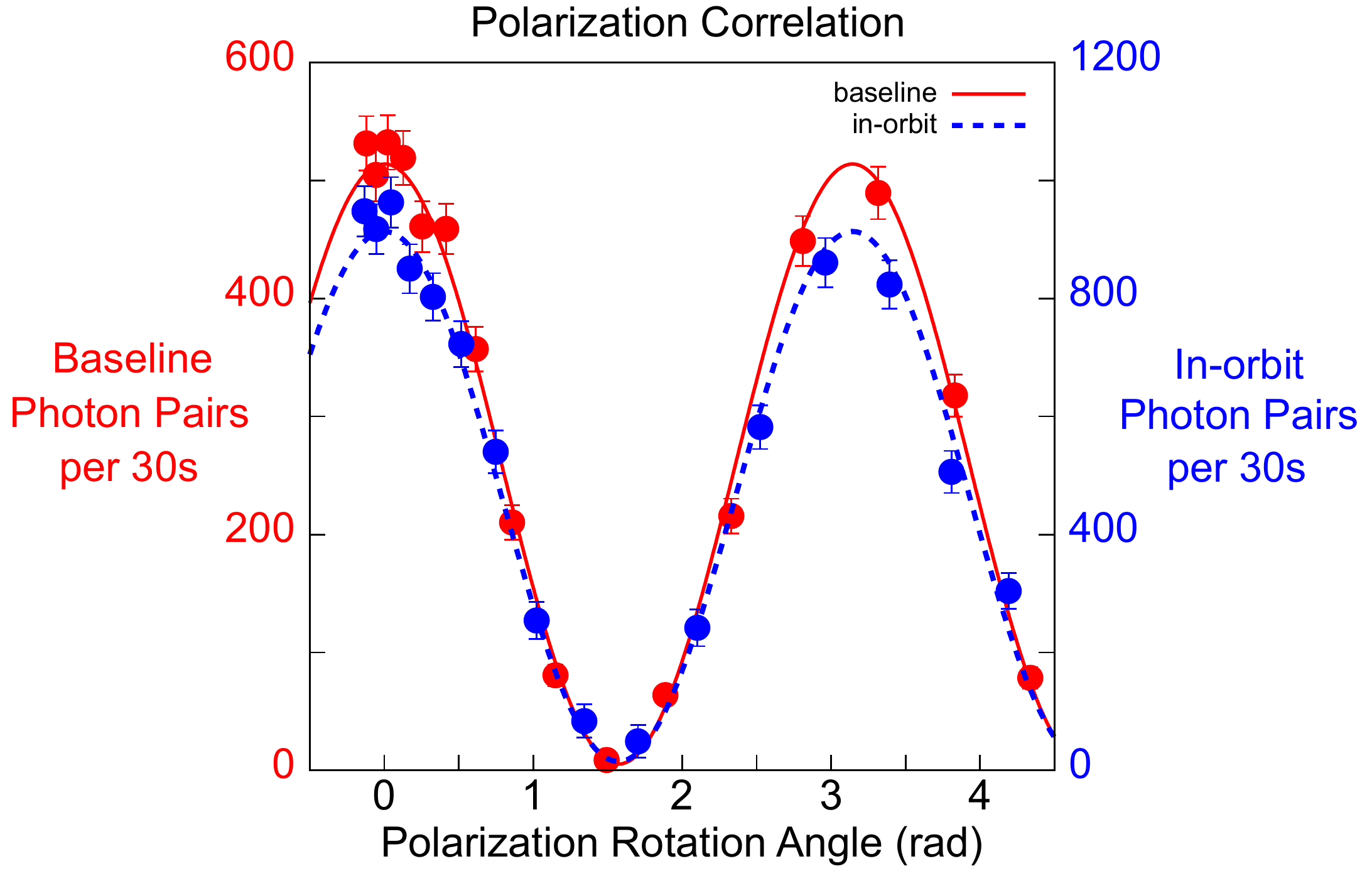}    
\caption{\label{fig:corr} Polarization correlation (after background subtraction) recorded in space and on Earth (integration time of \SI{30}{\s} at each setting). This data is used to generate the polarization rotator performance (lines) using Malus' law. The rotator achieves a polarization contrast of 97$\pm$2\% in both cases, and is sufficient for use in measuring the polarization entanglement between photon pairs.}
\end{figure}

The temperature experienced by the payload during the first experiment run in space is shown in Fig.~\ref{fig:data}~(a). Upon activation the payload box was heated to the target temperature of \SI{18}{\celsius}, after which the opto-electronics were activated sequentially. The GM-APDs were enabled first to measure their noise values (Fig.~\ref{fig:data}~(b)). Subsequently, the pump beam was enabled and stabilized at \SI{10}{\mW}, and collection of polarization correlation data started. The correlation data collected in this first experimental run is shown in Fig.~\ref{fig:corr}.

The in-orbit background rate for both detectors exceed the last (Earth-based) baseline data by approximately 30,000 events per second (Fig.~\ref{fig:data}~(b)). The observed increase in the background event rate is larger than predicted by simulation of asymptotic accumulation of radiation damage~\cite{tan13}. The cause for this increase is currently unknown and is under investigation. Despite this increase, the detectors are still operating in the linear regime. The closed-loop control circuit on the payload actively compensates for the increased rate by adjusting the operating voltage upwards~\cite{chandrasekara15}. 
This circuit has been shown to extend the saturation-free range of GM-APDs up to 600,000 events per second. 
It is unnecessary to utilize more sophisticated models for estimating the accidental coincidences generated when the detectors are in the saturation regime~\cite{grieve15}.
 
The average in-orbit brightness of the photon pair source is 60 co-incident detection events per second, while the rate of single photon detection events at APD1 and APD2 are 97,000 and 79,000 events per second, which is on the same order as the rate of dark counts. This brightness matches the results expected from ground-based tests conducted in the same temperature range. We note that the final baseline brightness shown in Fig.~\ref{fig:corr} is lower by almost a factor of two in comparison to the in-orbit data, due to the elevated temperature of the baseline test environment.  To avoid these temperature dependent effects, the flexure stage will be constructed using using materials having a lower coefficient of thermal expansion (e.g. titanium).
The performance of the polarization rotator is independently obtained from the observed polarization contrast in the baseline and in-orbit data (lines in Fig.~\ref{fig:corr}). The rotator achieves a contrast of (97$\pm$2\%) for both baseline and in-orbit performance, and has not been influenced so far by in-orbit radiation. 

The compatibility of the in-orbit polarization correlations with baseline measurements validates our design, and is clear evidence that the components of both source and polarization measurement system have experienced minimal degradation after 30 days in orbit. The GM-APDs continue to operate within the linear regime despite an elevated rate of background events. If further operation confirms that this is radiation-induced damage, it may be mitigated by improved shielding, e.g. a thicker aluminium housing. Further experiments will study the long-term performance of the payload, to be run at a rate of approximately twice per month (the spacecraft hosts two other experiments and power and data links need to be shared) and the results will be used to inform the design of the next iteration of our nanosatellite program.

\section*{Outlook}
\label{future}
The immediate objective of future work will be the extension of the existing SPDC source into an entangled photon system with sufficient brightness to overcome the anticipated link-loss in a space-to-ground optical link. Currently the correlated photon pairs use only a single BBO crystal and five additional crystals are needed to generate polarization entanglement by ensuring that the photon pairs are in a suitable superposition of polarization states. This is straightforward using the current design, as the crystals can be mounted on flexure stages and placed into pre-machined slots already available within the existing optical box. 

A greater challenge is to ensure that the entangled photon pair source exhibits high brightness and efficiency at the same time. This requires the use of beam shaping lenses, not included in the current design because of volume restrictions imposed by a multi-use nanosatellite. By moving to a dedicated nanosatellite of the same size (\SI{10}{\cm}~x~\SI{10}{\cm}~x~\SI{20}{\cm}) as the current spacecraft for a follow-up mission, there will be sufficient payload volume to enable beam shaping lenses. A recently concluded systematic study on the trade-off between beam size, crystal length and brightness~\cite{septriani15} has indicated that the SPDC geometry can be built to achieve a pair-to-singles efficiency of 30\%, while achieving brightness improvement by 3 orders of magnitude using the same pump power settings. These results will inform the final design.

This successful demonstration of an SPDC-based photon counting experiment on an orbiting nanosatellite is a major experimental milestone towards a space-based global quantum network. The challenge of demonstrating in-orbit performance for the components of future entangled photon experiments has been overcome. The iterative approach is also proving to be valuable, enabling the exact in-orbit performance of the components to be studied. For example, extensive ground testing and simulation failed to identify the anomalous increase in the APD dark count rate. The in-orbit data suggests that the space radiation damage model may need to be adjusted to account for previously unforeseen effects.  

Nanosatellites are a useful tool for quickly developing and demonstrating new miniaturised space technologies driven by their rapidly evolving capabilities~\cite{frost15}. Progress in the development of miniaturized attitude control systems and active pointing optics may make it feasible for very small spacecraft to carry out entanglement distribution experiments. More generally, we expect nanosatellites to play an increasingly important role in space-based quantum technology and science, as well as in other scientific endeavors that require cost-effective access to space.


\section*{Acknowledgements}
\label{ack}
The team from CQT thank Yau Y. S. for assistance with the initial design on the optical housing. The authors thank E.~Truong-Cao and H.~Askari for assitance with graphics. Valuable discussion on the manuscript was also held with C.~Kurtsiefer, J.~A.~Hogan, J.~A.~Grieve and R.~Bedington. The Galassia team acknowledges the support from Singapore Technologies Electronics in securing a launch opportunity. D.~K.~L.~Oi acknowledges the Scottish Quantum Information Network (QUISCO) and the EU FP7 CONNECT2SEA project \emph{``Development of Quantum Technologies for Space Applications"}. This research is supported by the National Research Foundation, Prime
Minister's Office, Singapore under its Competitive Research Programme (CRP
Award No. NRF-CRP12-2013-02, \emph{``Space based Quantum Key Distribution''}) and administered by the National University of Singapore. This program was also supported by the Ministry of Education, Singapore.

\section*{Methods}
\label{Methods}
\subsection*{The nanosatellite and payload operation}
\label{galassia}
\begin{figure}[ht]
\center
\includegraphics[width=0.75\linewidth]{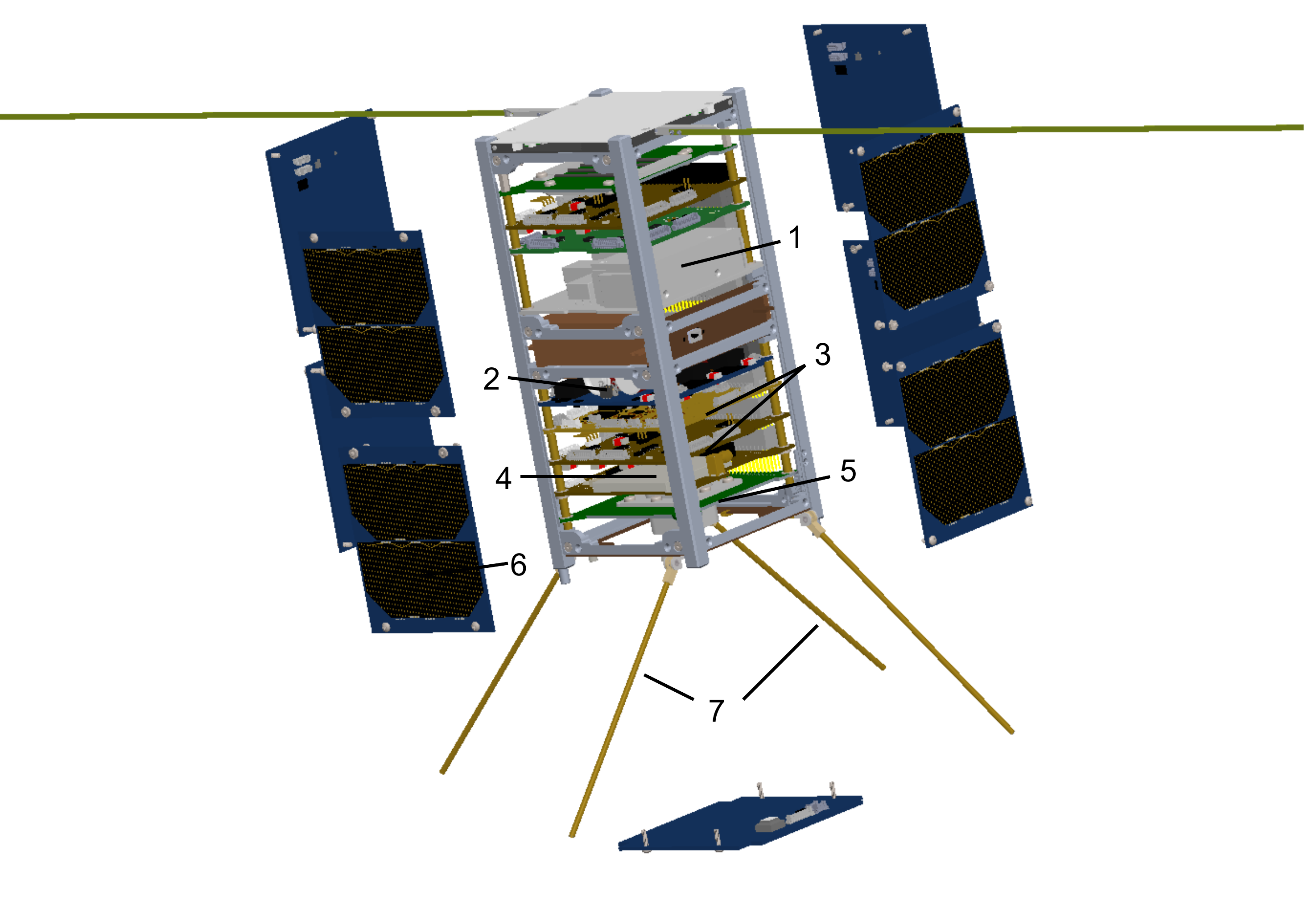}    
\caption{\label{fig:chassis} Exploded 3D model of the Galassia spacecraft showing optics payload (1), batteries (2), on-board computer (3), radio (4), passive attitude stabilization magnets (5), solar panels (6) and communication antennae (7).}
\end{figure}

The Galassia spacecraft is a satellite that is approximately \SI{20}{\cm} long with a \SI{10}{\cm} cross-section (see Fig.~\ref{fig:chassis}). The spacecraft is equipped with solar panels and a battery. It is capable of communicating with ground control via UHF radio link at a bit rate of approximately 1.2~kbps. The spacecraft is carrying two payloads one of which is the photon pair source described below. With both payloads, the spacecraft is approximately \SI{1.65}{\kg}. The spacecraft has an in-orbit lifetime of approximately 5 years before falling back to Earth.

The payload volume available is approximately \SI{9}{\cm}$\times$\SI{9}{\cm}$\times$\SI{4}{\cm}. 
The optics box is integrated with a custom-made printed circuit board~\cite{cheng15} to form a complete payload. The final payload mass was \SI{220}{\gram} with dimensions of approximately $95 \text{~mm} \times 90\text{~mm} \times 35\text{~mm}$. The circuit board serves two functions: to operate the optics experiment, and to provide a mechanical interface to the satellite. The SPDC source was pumped with an grating-stabilized diode laser capable of \SI{40}{\mW} optical output, but during space operation, the laser power is limited to \SI{10}{\mW}. The polarization rotators were custom devices and the control electronics for the experiment is built around an 8051 micro-controller.

A single experiment run consists of 16 voltage settings for the rotator. The data collection interval for each setting is \SI{1}{\s}, while the settling time was fixed at \SI{0.5}{\s} per setting. This enables a single experiment run to be completed in \SI{24}{\s}. For the in-orbit demonstration, experiments are run continuously and the results from the different runs are then integrated together. This integration increases signal-to-noise ratio, and enables the average performance of the rotator over the experiment time (during which temperature is expected to drift) to be assessed. It also enables the data collection to be robust against the occasional power fluctuations (caused by mode-hopping) that are experienced by the pump laser, during which time the experiment cannot operate.

While in orbit, the optical payload is designed to draw a peak power of \SI{2.5}{\W} for 10 minutes and \SI{1.3}{\W} for an additional 20 minutes. At the end of payload operation, a 64 kB data file containing the experiment data is placed in satellite's data storage unit to await transfer to ground control.
The spacecraft was launched on the PSLV C29 mission into an ``equatorial" orbit on Dec 16, 2015. 

\subsection*{Experiment profiles}
\label{profiles}

In order to acquire data under various scenarios and also have the ability to diagnose different components in the payload, 15 experimental profiles were uploaded. In Table. 1 we have outlined these profiles along with key parameters used to define them. In general each profiles will go through a heating phase (where heaters will be activated until turn-on temperature is reached), dark count phase (where dark counts will be collected) and experiment phase (where science data will be collected). The pump laser diode in all except three profiles (0$\times$37,0$\times$39, 0$\times$3A) is operated either with a fixed injection current or a fixed optical power (monitored by a photodiode on-board (PD)). The three profiles without pump activation serve solely to monitor dark counts (0$\times$37) and memory (0$\times$39, 0$\times$3A) respectively.

\begin{table}[ht]
\centering
\begin{tabular}{|c|c|c|c|c|c|c|}
\hline
ID   & \begin{tabular}[c]{@{}c@{}}Heating\\ (min)\end{tabular} & \begin{tabular}[c]{@{}c@{}}Dark\\ count\\ (min)\end{tabular} & \begin{tabular}[c]{@{}c@{}}Expt\\ time\\ (min)\end{tabular} & \begin{tabular}[c]{@{}c@{}}Current/\\ PD setting\end{tabular} & \begin{tabular}[c]{@{}c@{}}Memory\\ type\end{tabular} & \begin{tabular}[c]{@{}c@{}}Turn-on\\ temperature\\ ($^{\circ}$C)\end{tabular} \\ \hline
0x10 & 10                                                      & 3                                                            & 18                                                          & 10mW                                                          & Flash                                                 & 18.7                                                                \\ \hline
0x30 & 10                                                      & 3                                                            & 18                                                          & 10mW                                                          & Flash                                                 & 12.4                                                                \\ \hline
0x31 & 10                                                      & 3                                                            & 18                                                          & 10mW                                                          & Flash                                                 & 23.6                                                                \\ \hline
0x32 & 10                                                      & 3                                                            & 18                                                          & 7mW                                                           & Flash                                                 & 18.7                                                                \\ \hline
0x33$^{\ast}$ & 10                                                      & 3                                                            & 18                                                          & 10mW                                                          & Flash                                                 & 18.7                                                                \\ \hline
0x34 & 5                                                       & 3                                                            & 8                                                           & 7mW                                                           & Flash                                                 & 12.4                                                                \\ \hline
0x35 & 10                                                      & 3                                                            & 18                                                          & 37mA                                                          & Flash                                                 & 18.7                                                                \\ \hline
0x36 & 10                                                      & 3                                                            & 18                                                          & 34mA                                                          & Flash                                                 & 18.7                                                                \\ \hline
0x37 & 0                                                       & 23                                                           & 0                                                           & NA                                                            & Flash                                                 & NA                                                                  \\ \hline
0x38$^{\dagger}$ & 10                                                      & 3                                                            & 18                                                          & 37mA                                                          & Flash                                                 & 18.7                                                                \\ \hline
0x39 & 0                                                       & 0                                                            & 0                                                           & NA                                                            & Flash                                                 & NA                                                                  \\ \hline
0x3A & 0                                                       & 0                                                            & 0                                                           & NA                                                            & EEPROM                                                & NA                                                                  \\ \hline
0x3B & 10                                                      & 3                                                            & 15                                                          & 10mW                                                          & EEPROM                                                & 18.7                                                                \\ \hline
0x3C & 10                                                      & 3                                                            & 15                                                          & 37mA                                                          & EEPROM                                                & 18.7                                                                \\ \hline
0x3D & 10                                                      & 3                                                            & 5                                                           & 37mA                                                          & Flash                                                 & 12.4                                                                \\ \hline
\end{tabular}
\caption{Fifteen profiles are designed for the mission each with a different purpose. The $0\times33(^{\ast})$ profile uses a different high voltage on the APD in anticipation of radiation damage. The $0\times38(^{\dagger})$ profile differs from the others because it rotates idler (867nm) instead of signal (760nm).}
\label{exptprofile}
\end{table}

\end{document}